\documentclass[twocolumn,english,aip,sd,amsmath,amssymb,reprint]{revtex4-1}
\usepackage[T1]{fontenc}
\usepackage[latin9]{inputenc}
\usepackage{babel}
\usepackage{textcomp}
\usepackage{amsmath}
\usepackage{amssymb}
\usepackage{graphicx}
\usepackage{subscript}
\usepackage[unicode=true,pdfusetitle,
 bookmarks=true,bookmarksnumbered=false,bookmarksopen=false,
 breaklinks=false,pdfborder={0 0 1},backref=false,colorlinks=false]
 {hyperref}

\makeatletter

\pdfpageheight\paperheight
\pdfpagewidth\paperwidth

\makeatother

\begin{document}

\title{Lithium ion intercalation in thin crystals of hexagonal TaSe\textsubscript{2} gated by a polymer electrolyte}

\author{Yueshen Wu}
\affiliation{Key Laboratory of Artificial Structures and Quantum Control and Shanghai
Center for Complex Physics, School of Physics and Astronomy, Shanghai
Jiao Tong University, Shanghai 200240, China}

\author{Hailong Lian}
\affiliation{Key Laboratory of Artificial Structures and Quantum Control and Shanghai
Center for Complex Physics, School of Physics and Astronomy, Shanghai
Jiao Tong University, Shanghai 200240, China}

\author{Jiaming He}
\affiliation{Key Laboratory of Artificial Structures and Quantum Control and Shanghai
Center for Complex Physics, School of Physics and Astronomy, Shanghai
Jiao Tong University, Shanghai 200240, China}

\author{Jinyu Liu}
\affiliation{Department of Physics and Engineering Physics, Tulane University,
New Orleans, LA 70118, USA}

\author{Shun Wang}
\altaffiliation{Current Affiliation: MOE Key Laboratory of Fundamental Physical Quantities Measurements, School of Physics, Huazhong University of Science and Technology, Wuhan 430074, China}
\affiliation{Key Laboratory of Artificial Structures and Quantum Control and Shanghai
Center for Complex Physics, School of Physics and Astronomy, Shanghai
Jiao Tong University, Shanghai 200240, China}

\author{Hui Xing}
\email{huixing@sjtu.edu.cn}
\affiliation{Key Laboratory of Artificial Structures and Quantum Control and Shanghai
Center for Complex Physics, School of Physics and Astronomy, Shanghai
Jiao Tong University, Shanghai 200240, China}

\author{Zhiqiang Mao}
\affiliation{Department of Physics and Engineering Physics, Tulane University,
New Orleans, LA 70118, USA}

\author{Ying Liu}
\email{yxl15@psu.edu}
\affiliation{Key Laboratory of Artificial Structures and Quantum Control and Shanghai
Center for Complex Physics, School of Physics and Astronomy, Shanghai
Jiao Tong University, Shanghai 200240, China}
\affiliation{Department of Physics and Materials Research Institute, Pennsylvania
State University, University Park, PA 16802, USA}
\affiliation{Collaborative Innovation Center of Advanced Microstructures, Nanjing
210093, China}

\date{\today}
\begin{abstract}
Ionic liquid gating has been used to modify properties of layered transition metal dichalcogenides (TMDCs), including two-dimensional (2D) crystals of TMDCs used extensively recently in the device work, which has led to observations of properties not seen in the bulk. The main effect comes from the electrostatic gating due to strong electric field at the interface. In addition, ionic liquid gating also leads to ion intercalation when the ion size of gate electrolyte is small compared to the interlayer spacing of TMDCs. However, the microscopic processes of ion intercalation have rarely been explored in layered TMDCs. Here, we employed a technique combining photolithography device fabrication and electrical transport measurements on the thin crystals of hexagonal TaSe\textsubscript{2} using multiple channel devices gated by a polymer electrolyte LiClO$_{4}$/PEO. The gate voltage and time dependent source-drain resistances of these thin crystals were used to obtain information on the intercalation process, the effect of ion intercalation, and the correlation between the ion occupation of allowed interstitial sites and the device characteristics. We found a gate voltage controlled modulation of the charge density waves and scattering rate of charge carriers. Our work suggests that ion intercalation can be a useful tool for layered materials engineering and 2D crystal device design.
\end{abstract}

\maketitle
Atomically thin crystals of layered transition metal dichalcogenides (TMDCs) were found to show physical properties including basic band structure that are drastically different from that of the bulk\citep{PRL.108.196802,RN512}, making mechanical exfoliation preparation of such atomically thin crystals a powerful route for materials discovery. Ionic gating was used to tune the properties of these interesting materials further, leading to remarkable success.\citep{RN119-TaS2iFET,RN247,RN365-int,PRBFeSe,RN196-NbSe2iFET} By supplying a exceedingly large number of charge carriers, the physical properties of these layered compounds were shown to be greatly tunable. The ionic gating involves both electrostatic tuning of the surface charge carriers through an electric field and ionic intercalation of the layered compound in the presence of gate electrolyte with small ion size. The latter possibly modifies the interlayer coupling in these few-atomic-layer TMDCs. While its ability to manipulate the properties of these atomically thin crystals were well recognized,\citep{RN119-TaS2iFET,RN247} the underlying physical processes of ion intercalation were not fully resolved. Understanding how ion intercalate in atomically thin layered TMDCs, including the kinetics, stages, interstitial sites of occupation, and the distribution of the ions, is important for materials engineering and the designing of functional electronic devices.

As a form of tantalum diselenide, 2H-TaSe$_2$ features a hexagonal crystal structure with a stack of one Ta layer sandwiched between two Se layers in an ABAB stacking. 2H-TaSe$_2$ is a superconductor with an intrinsic transition temperature of 0.15 K.\citep{RN357-SC,RN358-SC,RN356-SC} It also features an incommensurate charge density wave (ICDW) transition at 120 K followed by a commensurate charge density wave (CCDW) transition at 90 K.\citep{RN363-CDW,RN361-CDW,RN346-CDW,RN348-CDW,RN349-CDW,RN216-CDW} In the bulk form, intercalating 2H-TaSe$_2$ by various ions, such as Fe, Ni or Pd, into 2H-TaSe\textsubscript{2} was found to lead to changes in both superconducting and CDW properties of this compound.\citep{RN353-int,RN360-int,RN354-int,RN355-int,RN352-int,RN203-int} X-ray diffraction (XRD)\citep{RN344-int} and M\"{o}ssbauer spectroscopy\citep{RN347-int} studies revealed that, for nearly stoichiometric bulk LiTaSe\textsubscript{2}, Li ions occupied octahedral sites between TaSe\textsubscript{2} layers with the valence of the Ta changing from Ta\textsuperscript{4+} to Ta\textsuperscript{3+}, suggesting that each Li ion contributes one electron to the host material. Optical transmission studies showed that the Li intercalated 2H-TaSe$_2$ is semiconducting, consistent with the results of theoretical calculations.\citep{RN360-int,Lical}

Atomically thin single crystals of 2H-TaSe\textsubscript{2} were found to exhibit properties not seen in the bulk. For example, weak antilocalization was observed and attributed to a strong spin-orbital coupling.\citep{RN301-2D} Thermal conductivity was seen to be suppressed from that in the bulk.\citep{RN288-2D} In all these studies, however, the process of Li ion intercalation and its effect on the electrical transport in the thin crystal were not studied in detail. In this work, we attempted to address these issues by employing a technique combining photolithography device fabrication and electrical transport measurements on thin crystals of 2H-TaSe\textsubscript{2} featuring multiple channels gated by a polymer electrolyte.

\begin{figure}
\includegraphics[width=0.8\linewidth]{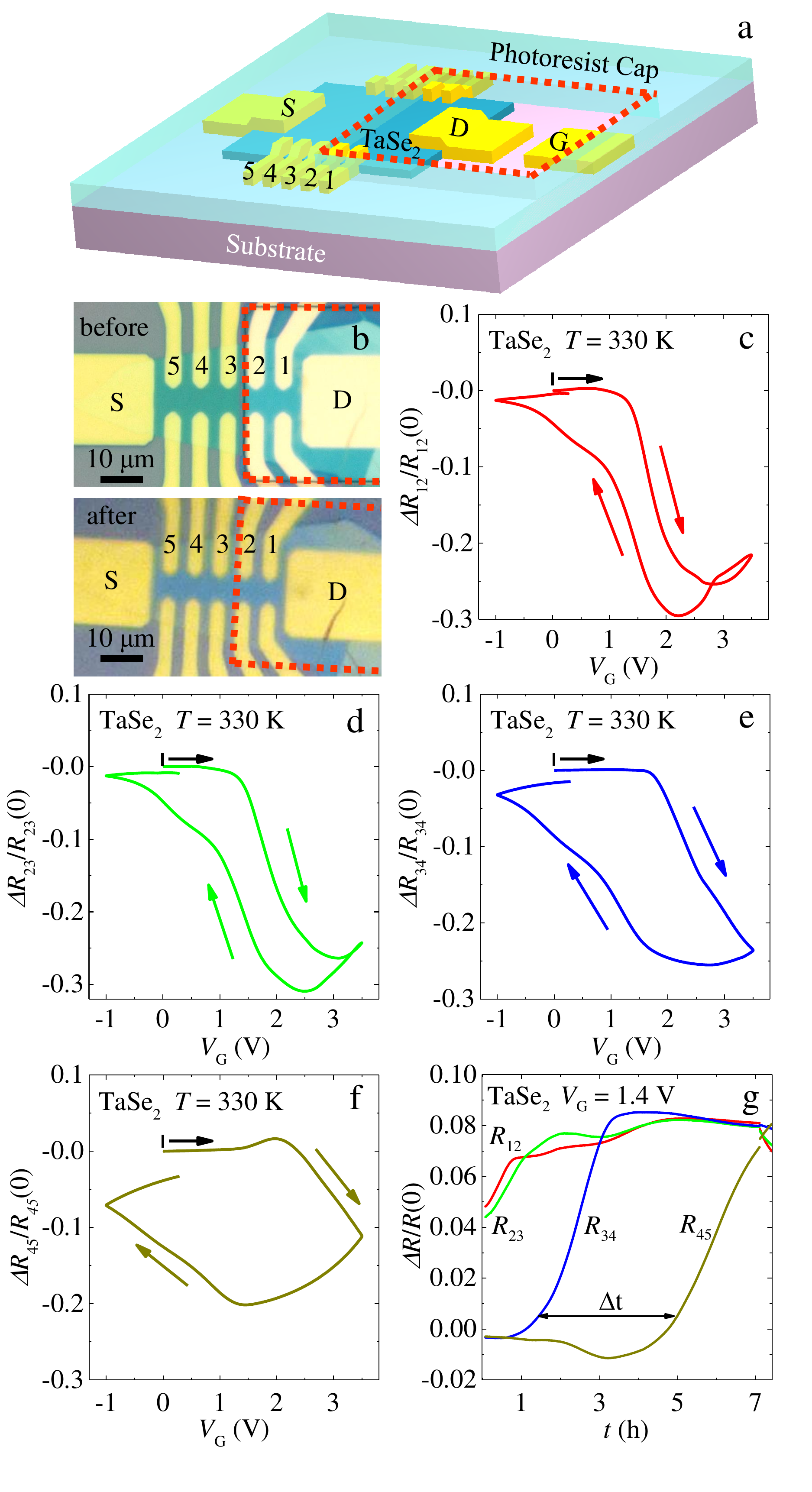}\caption{\label{fig1}(a) Schematic of multi-channel field-effect transistor device gated by a polymer electrolyte. The photoresist covers the part of the crystal. (b) Optical image of the device prepared using a 10-nm-thick 2H-TaSe$_2$ single crystal before (upper panel) and after (lower panel) ionic gating. The photoresist window is outlined by the dash line. (c-f) Values of $R_{12}$, $R_{23}$, $R_{34}$, and $R_{45}$ as a function of $V$\protect\textsubscript{G} (see text). (g) Time dependent resistance changes of $R_{12}$, $R_{23}$, $R_{34}$, $R_{45}$ at $V$\protect\textsubscript{G}$=1.4$ V, respectively.}
\end{figure}

Bulk single crystals of 2H-TaSe\textsubscript{2} were grown by a chemical vapor transport method. Atomically thin crystals of 2H-TaSe$_2$ were obtained by mechanical exfoliation and transferred onto a 300-nm-thick SiO\textsubscript{2}/Si substrate by PDMS stamp.\citep{2053-1583-1-1-011002} The crystal thickness was estimated by a color code that infers the thickness of the crystal based on the color and faintness of the color calibrated by atomic force microscope. Device pattern was defined by photolithography, with the contacts prepared by the deposition of a 10-nm-thick Ti and 100-nm-thick gold film in series, followed by a lift-off process. To probe the effect of ion intercalation on the electronic properties of the atomically thin crystals of TaSe\textsubscript{2}, it is useful to distinguish the contribution of the electric field gating and ion intercalation on the sample. For electrostatic gating, Li ions form electron double layers at the surface, which affects the channel resistance through only the top layer of the crystal due to electrostatic screening of a metallic sample. However, the intercalation of Li ions, which in our devices enter the crystal from the edges and are inserted between the atomic layers, affects the properties of the entire crystal. We developed a technique to study the ion intercalation process by directing ions to enter from selected part of an atomically thin crystal and monitor the ion diffusion in the crystal through the measurement of the resistances of spatially separated channels as a function of gate voltage and time. As shown in Fig. 1a and b, this technique features a photoresist window that covers only part of a 10-nm-thick crystal, allowing Li ions to enter the crystal from the uncovered sides.

The polymer electrolyte was prepared by mixing LiClO\textsubscript{4}/PEO, 0.3 g and 1 g, respectively, which was dissolved in 15 ml anhydrous methanol, followed by stirring for 10 hours at the temperature of 50 \textcelsius . A droplet of the electrolyte was applied on the device covering both channel and gate leads. The entire sample was baked at 97 \textcelsius \  for an hour to remove residual solvent in high vacuum (10\textsuperscript{-5} Torr) just before measurements. We ramped the gate voltage ($V$\textsubscript{G}) at 330 K with a constant sweeping rate at 0.5 mV/s. The gate voltage dependences of the channel resistance between neighboring voltage leads, $R_{12}$, $R_{23}$, $R_{34}$, $R_{45}$, shown in Figs.1c-f, were simultaneously measured as the gate voltage was swept. Previous studies show that the ions cease to diffuse in this polymer electrolyte below 280 K.\citep{PEO,PEO1998} For the channel resistance as a function of temperature measurements, a wait time of 90-min was used to allow the ions to diffuse into the crystal before the device was cooled down to lower temperatures. After all the measurements, no obvious chemical reaction was observed in our sample as shown in the lower panel of Fig. 1b.

Gate voltage dependence of resistance for source-drain channels of fully covered parts of the crystal, $R_{34}$ (Fig. 1e) and $R_{45}$ (Fig. 1f) as well as that of the uncovered part  $R_{12}$ (Fig. 1c) were obtained at 330 K. Interestingly, the covered and uncovered parts of the crystal were found to respond to the gate voltage similarly, which indicates that the ion intercalation plays a dominant role in determining the channel resistance. At low gate voltage in uncovered part, the resistance increased slightly with increasing gate voltage, suggesting a weak intercalation effect mixed with electrostatic gating. In this regime, it is difficult to determine the weight of both effects since the large carrier density in metallic TaSe\textsubscript{2} makes electrostatic gating effect much less pronounced than that found in semiconducting MoS\textsubscript{2}\citep{YeMoS2}. It was also found that the hysteresis in the $R$ $vs.$ $V$\textsubscript{G} curves was larger in the covered channels than that of the uncovered ones, indicating that the Li diffusion in the crystal is significant even during gate voltage sweep. Very recently, ultrafast Li diffusion in bilayer graphene was found,\citep{RN513} which suggests that transport measurements may be used to study the ion diffusion in a 2D crystal. In the present work, the channel resistances of the spatially separated pairs of covered part of the device were seen to start to increase in sequence (Fig. 1g). If the onset of resistance increase was used to mark the start of the Li intercalation, and the covered part of the device can be approximated as a one-dimensional diffusion model, we can estimate the diffusion coefficient of the Li ion at the starting of intercalation using the equation $D=\frac{\left(x_{4}^{2}-x_{3}^{2}\right)}{2\Delta t}$, as done previously.\citep{reif2009fundamentals,InTaSe2}  Here the length $x_{i}$ is the distance of \textit{i}-th voltage lead away from photoresist window. It appears that Li ions took roughly $\Delta t$ = 3.5 hours to diffuse from voltage leads 3 to 4, so that $D$ is approximately $2.8\text{\texttimes}10^{-11}cm^{2}/s$ at 330 K. This shows that the diffusion constant $D$ can be measured in our device using the electrical transport measurements.  However, it is important to note that $D$ is affected by many factors, including Li concentration, $V$\textsubscript{G} and the chemical structure of the electrolyte,\citep{KUMAGAI1991935,doi:10.1021/ar200329r} making the precise determination of the diffusion constant difficult.

Physical properties of this thin crystal of 2H-TaSe$_2$ were found to evolve continuously as the Li ions intercalate it. In Fig. 2, the evolution of the temperature dependence of sample resistance of an uncovered sample taken at a fixed gate voltage is shown. During the measurements of $R$ $vs.$ $T$ curves, the gate voltage is ramped to a desired value at 330 K and the sample was cooled down from 330 to 2 K at a constant rate 1 K/min. The sample was warmed back up at the same rate to 330 K, with the channel resistance measured during the warming up as well. The values of the channel resistance obtained during the cooling down and warming up were found to differ only above a relatively high temperature, T $\approx$ 280 K, suggesting that the Li ions were frozen in below this temperature, consistent with previous studies.\citep{PEO,PEO1998} The $R$ $vs.$ $T$ curves, which were obtained as the sample is cooled down, revealed a continuous evolution of behavior as the gate voltage was varied. It is seen that, at the small gate voltages, the sample exhibit metallic behavior over the entire temperature range, with the $R$ $vs.$ $T$ curve shifting to higher value as $V$\textsubscript{G} increases. A hump was found in the $R$ $vs.$ $T$ curve around 120 K (Fig. 2a), which marks the CDW transition in this 10-nm thick crystal of TaSe\textsubscript{2}. It becomes weaker and eventually disappeared with increasing gate voltage, suggesting
the suppression of CDW by the Li ion intercalation. There are earlier efforts in modulating CDW by ionic liquid gating. For example, in TaS\textsubscript{2} electron double layer transistors, the transition temperature of CDW can be tuned from 190 K to 140 K due to the change of Fermi surface topology by carrier doping\citep{YeTaS2}. Other effect was found in Cu doped TaS\textsubscript{2}\citep{RN417} in which the disruption of CDW coherence is induced by a random distribution of ions. As the gate voltage increased further, the overall behavior of resistance were found to show a decrease overall with increasing $V$\textsubscript{G}, as shown in Fig. 2b. In addition, a crossing of the $R$ $vs.$ $T$ curves taken at different $V$\textsubscript{G} was found. Similar features were found in Li intercalated bulk crystals of 2H-TaS\textsubscript{2}.\citep{RN200LiTaS2} Both observations suggest that the processes are more complex than simple addition of charge carriers to the system. Finally, at the highest values of $V$\textsubscript{G}, the channel resistance was found to rise rapidly with increasing $V$\textsubscript{G}, with an upturn in channel resistance in the low-temperature limit for the highest $V$\textsubscript{G} (Fig. 2c). A resistance anomaly around 200 $\sim$ 250 K was also observed, above which the resistance slope $dR/dT$ increased slightly.

\begin{figure}
\includegraphics[width=1\linewidth]{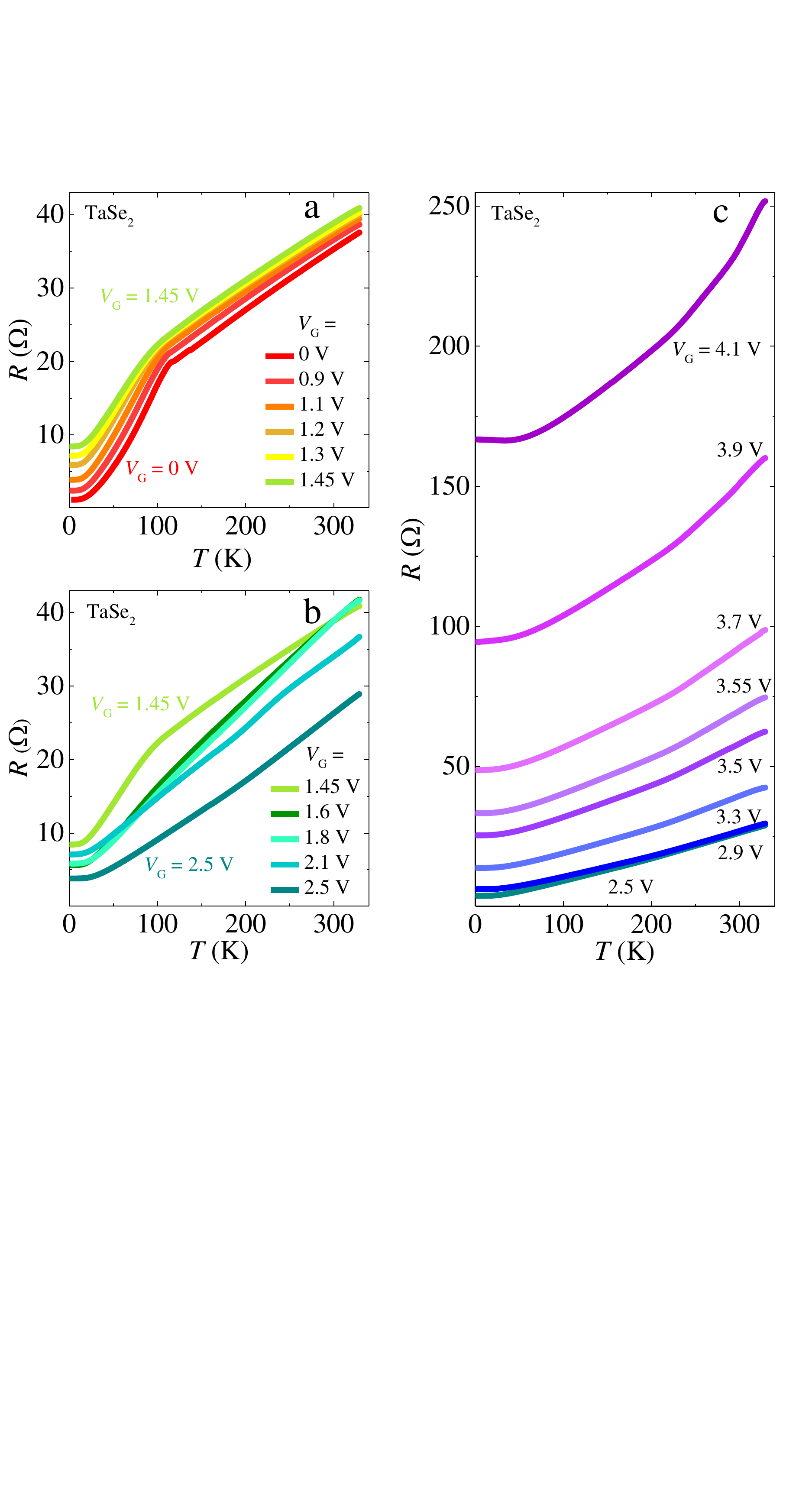}

\caption{\label{fig3}(a-c) $R$ $vs.$ $T$ curves with an up-sweep of $V$\protect\textsubscript{G} up to 4.1 V. The hump around 120 K signaling the onset of CDW was found to become weaker and eventually disappear at low gate voltage. A resistance anomaly around 200 $\sim$ 250 K was also observed at high gate voltages. }
\end{figure}

We propose that the evolution of the electronic transport properties revealed in Fig. 2 is linked to the distribution of the Li ions during the intercalation process. Upon entering the crystal, Li ions will adopt several possible sites between layers featuring different coordination environments. Based on the theoretical calculation and experimental studies in the bulk,\citep{RN344-int,Lical}  the most favorable intercalation sites between the layers are the octahedral sites, which will be taken up the first, as shown schematically in Fig. 3c. The next favorable ones are tetrahedral sites (Fig. 3e). Even at the initial process of the Li ion intercalation on the octahedral sites, ordered ion distribution may be formed due to ionic Coulomb repulsion (Fig. 3d). For Li ions occupying the octahedral sites at lowest values of $V$\textsubscript{G}, the electrons will tend to be bound to the positive charged ions, which will reduce the number of the hole carriers (electron doping). If these positively charged ions are randomly distributed, it will tend to enhance scattering of conducting carriers. Once ordered ion distribution occurs, however, the scattering of carriers will change. Finally, as tetrahedral sites started to be occupied, which appeared to contribute different scatterings of charge carriers, distinct features in $R$ $vs.$ $T$ including a negative $dR/dT$ at low temperatures for the highest $V$\textsubscript{G} and a bump near 200 - 250 K were found even though the precise origin of these features have not yet been identified.

\begin{figure}
\includegraphics[width=1\linewidth]{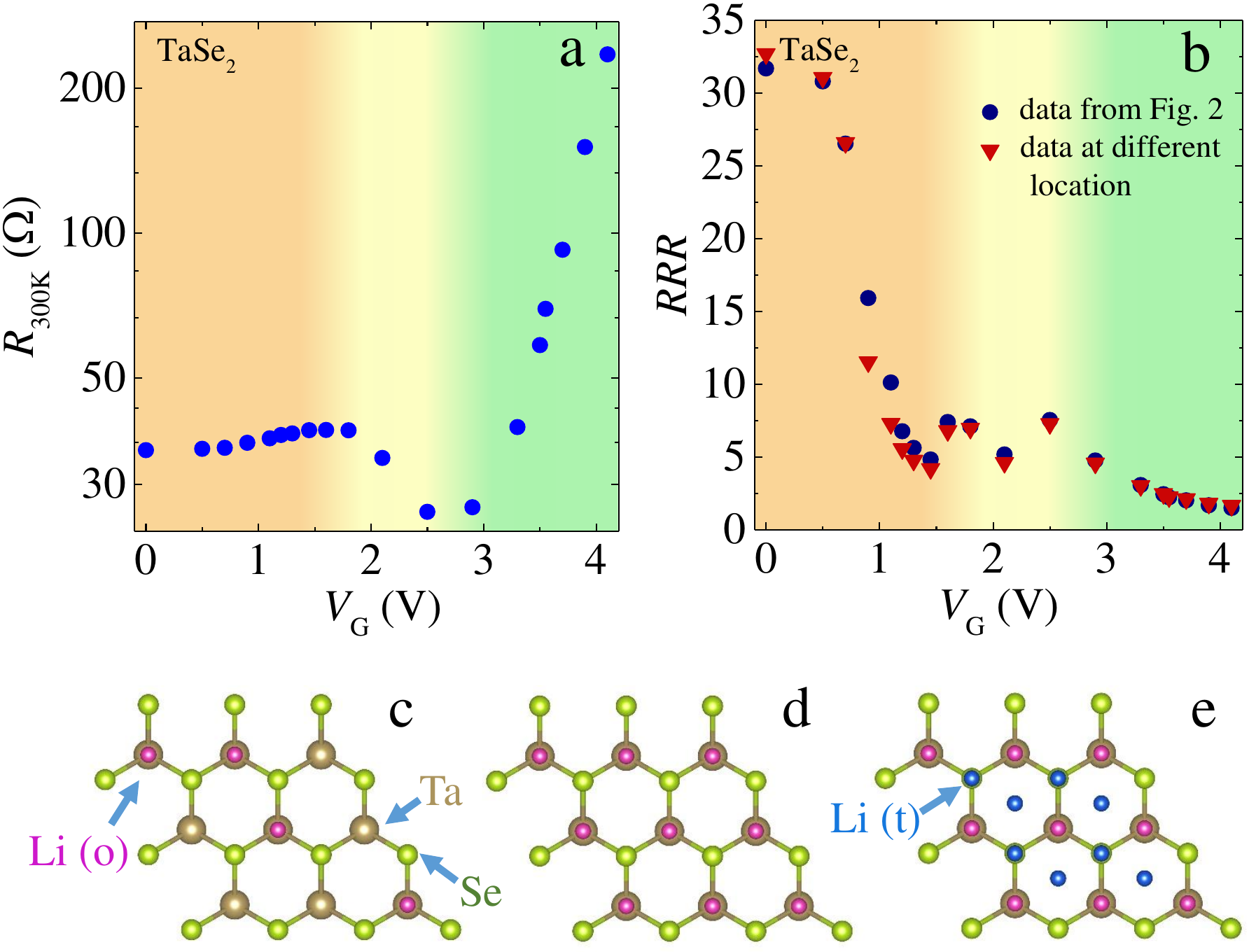}

\caption{\label{fig4}$V$\textsubscript{G} dependences of $R$\protect\textsubscript{300K} (a) and $RRR$ at different locations (b). The evolution of behavior is marked by different color and correlated with the ion distribution during the ion intercalation process. (c-d) Schematics of randomly distributed Li on octahedral (o) sites, an ordered structure of Li on octahedral (o) sites, and an ordered structure of Li on octahedral sites together with a randomly distributed Li on tetrahedral (t) sites.}
\end{figure}

Values of the room-temperature channel resistance, $R$\textsubscript{300K}, and residual resistance ratio, $RRR$ = $R$\textsubscript{330K}/$R$\textsubscript{2K} are plotted against $V$\textsubscript{G} in Figs. 3a and b. As $V$\textsubscript{G} increases, $R$\textsubscript{300K} is seen to first stay roughly a constant, increasing only slightly as $V$\textsubscript{G} increases. At $V$\textsubscript{G} = 1.6 V, $R$\textsubscript{300K} is seen to drop suddenly. Above $V$\textsubscript{G} = 2.5 V,  $R$\textsubscript{300K} was found to increase sharply as the gate voltage was increased further. Similar behavior was seen in the $RRR$ $vs.$ $V$\textsubscript{G} plot. Consistent behaviors of $RRR$ vs. gate voltage were found at different locations of the sample, with minor differences in detailed behavior of resistance and $RRR$, which could be due to unavoidable inhomogeneity of Li occupation. It is interesting to note that the suppression of CDW, the increase in $R$\textsubscript{300K}, and the decrease in $RRR$ appear to be correlated with the Li ion distribution proposed above. In particular, the randomly distributed Li ions in octahedral sites is transformed into an ordered structure when the concentration of Li ions is sufficiently high (Fig. 3d), as seen in other intercalated TMDCs with a crystal structure identical with 2H-TaSe\textsubscript{2} that feature staging and disorder-order transition.\citep{RN345-2H-TaS2} Decreasing $RRR$ at the highest $V$\textsubscript{G} values may indicate that the scattering from impurity is enhanced. It is likely that some Li ions are trapped in tetrahedral sites instead of octahedral sites, leading to enhanced scatterings of charge carriers (Fig. 3e). It should be noted that a previous study of bulk Pd\textsubscript{x}TaSe\textsubscript{2}\citep{RN203-int} showed that a CDW phase was found to be suppressed by ion intercalation. There $RRR$ was found to decrease monotonically as the ion concentration increased. The maximum doping level was 0.14 in Pd\textsubscript{x}TaSe\textsubscript{2}. No evidence for Pd ions to form an ordered structure in such concentration level was found. The increase in Pd ion concentration leads to a decreased $RRR$. These differences suggest that not only carrier doping but also scattering from ions affect the physical properties of thin crystals of TaSe\textsubscript{2}. It is likely that the Li concentration in our devices was higher than that of Pd in Pd\textsubscript{x}TaSe\textsubscript{2}. Unfortunately, it is difficult to estimate the Li concentration by Hall measurement in the current work due to multiband nature of TaSe\textsubscript{2}. More experiments are needed to confirm the picture proposed above.

In summary, we developed a technique to study Li intercalation and its effect on physical properties of atomically thin crystals of 2H-TaSe\textsubscript{2} by electrical transport measurements using polymer electrolyte gated 2D-crystal devices featuring covered and uncovered channels. We found that the Li ion diffusion coefficient is $2.8\text{\texttimes}10^{-11}cm^{2}/s$ at 330 K. The effect of ion intercalation on the properties of the thin crystals of TaSe$_2$ and possible correlation between the ion occupation of allowed interstitial sites and the device characteristics were demonstrated. A gate voltage controlled modulation of the charge density waves and scattering rate of charge carriers established in this work may help advance the practical use of 2D crystal materials.

We acknowledge useful discussion with Dr. Zhe Wang. The work done in China is supported by MOST of China (Grant Nos. 2015CB921104 and 2014CB921201), National Natural Science Foundation of China (Grant Nos. 11474198, 11521404, and 91421304), the CAS/SAFEA international partnership program for creative research teams of China. Work at Penn State is supported by NSF under Grant No. EFMA1433378 and at Tulane supported by the U.S. Department of Energy under EPSCoR Grant No. DESC0012432 with additional support from the Louisiana Board of Regents.

\bibliographystyle{apsrev4-1}

\end{document}